# Anisotropic Acoustic Plasmons in Black Phosphorus

In-Ho Lee[1], Luis Martin-Moreno[2,*], Daniel A. Mohr[1], Kaveh Khaliji[1],

Tony Low[1,*], and Sang-Hyun Oh[1,*]

[1] Department of Electrical and Computer Engineering, University of Minnesota, Minneapolis, Minnesota, 55455, U.S.A.

[2] Instituto de Ciencia de Materiales de Aragón and Departamento de Física de la Materia Condensada, CSIC-Universidad de Zaragoza, E-50009 Zaragoza, Spain.

*E-mail: lmm@unizar.es (L.M.-M.), tlow@umn.edu (T.L.), sang@umn.edu (S.-H.O.)

Dated: December 29, 2017

**TABLE OF CONTENTS GRAPHIC**

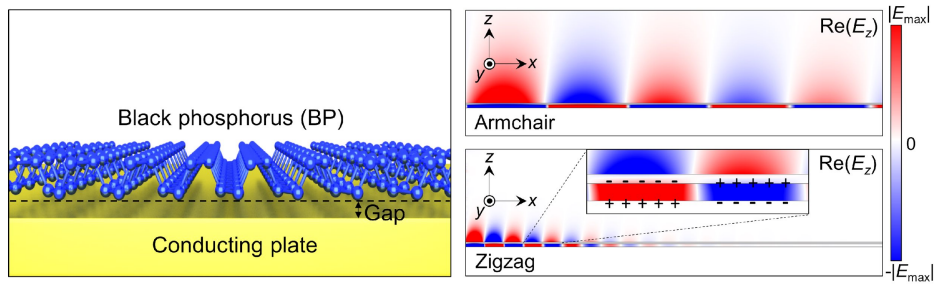

**ABSTRACT**

Recently, it was demonstrated that a graphene/dielectric/metal configuration can support 'acoustic plasmons', which exhibit extreme plasmon confinement an order of magnitude higher than that of conventional graphene plasmons. Here, we investigate acoustic plasmons supported in a monolayer and multilayers of black phosphorus (BP) placed just a few nanometers above a conducting plate. In the presence of a conducting plate, the acoustic plasmon dispersion for the armchair direction is found to exhibit the characteristic linear scaling in the mid- and far-infrared regime while it largely deviates from that in the long-wavelength limit and near-infrared regime. For the zigzag direction, such scaling behavior is not evident due to relatively tighter plasmon confinement. Further, we demonstrate a new design for an acoustic plasmon resonator that exhibits higher plasmon confinement and resonance efficiency than BP ribbon resonators in the mid-infrared and longer wavelength regime. Theoretical framework and new resonator design studied here provide a practical route toward the experimental verification of the acoustic plasmons in BP and open up the possibility to develop novel plasmonic and optoelectronic devices that can leverage its strong in-plane anisotropy and thickness-dependent band gap.

Two dimensional (2D) materials[1, 2] have attracted enormous interest due to their unique properties such as ultrahigh charge carrier mobility,[3-5] anomalous quantum Hall effect,[6, 7] and strong light-matter interaction.[8, 9] Among a variety of such exciting properties, strong light-matter interactions in 2D materials are particularly intriguing considering the extreme size mismatch between their atomic-scale thicknesses and wavelengths of free-space light. Moreover, this feature plays a central role in many potential applications of 2D materials such as optical modulators,[10, 11] metasurfaces,[12, 13] biosensors,[14-16] and photodetectors.[17-20] For a particular set of 2D materials including graphene, light-matter interactions can be even more intense because of the excitation of surface plasmons.[21-23] Compared to conventional surface plasmons in noble metals, the plasmons in 2D materials exhibit tighter confinement[24, 25] ($\sim\lambda_0/100$) as well as tunability by extrinsic doping.[22, 26] Many researchers have demonstrated that such features allow for the development of nanoscale photonic and optoelectronic devices that have novel functionality and superior performance inaccessible with conventional materials.[15, 27, 28]

Recent work on graphene indicates that the plasmon wavelength, and accordingly the confinement of 2D plasmons, can be further reduced in the presence of a conducting plate adjacent to the graphene.[29-32] As in the case of spatially separated double-layer graphene,[33] the hybridization of two plasmons in a graphene sheet and its mirror image result in less confined 'optical' and highly confined 'acoustic' plasmon modes. Since the charges oscillate out-of-phase, the vertical electric fields of acoustic plasmons are largely confined within the nanometric gap with a conducting plate, which gives extreme plasmon confinement defined by the gap size. In contrast to conventional graphene plasmons or 2D electron gas that have a parabolic dispersion, interestingly, acoustic plasmons exhibit a linear dispersion at small frequencies.[30]

Recently, black phosphorus (BP) has been extensively studied as a novel anisotropic plasmonic material. In contrast to other 2D plasmonic materials, the inherent in-plane anisotropy of BP renders the plasmon dispersion dependent on the propagation direction.[34] These anisotropic plasmons are expected to enable the development of novel polarization-dependent optoelectronic devices such as optical modulators,[35, 36] tunable polarization rotators,[37, 38] and polarization-sensitive photodetectors.[39, 40] One possible way to maximize the light-matter interaction in BP for such applications is to leverage the extreme confinement of acoustic plasmons. In this regard, it is imperative to understand how the in-plane anisotropy of BP is manifested through the acoustic plasmon dispersion and how these plasmons enhance the light-matter interactions in BP. For practical applications, in addition, new resonator configurations for acoustic plasmons that require minimal post-processing after the BP deposition should be investigated due to its instability in the ambient environment.[41, 42]

In this work, we theoretically investigate the dispersion of acoustic plasmons in freestanding BP placed adjacent to a conducting plate. Using both analytical and numerical approaches, we study how the in-plane anisotropy of BP is reflected in the plasmon dispersion and how the acoustic plasmons scale with frequency $\omega$ and gap size $g$ with a conducting plate. The effect of doping and BP thickness is examined as well. Further, we propose a practically viable and highly efficient design for an acoustic plasmon resonator. A modified Fabry-Perot resonance model is developed to describe the resonant behavior of such an acoustic plasmon resonator.

In the geometrical configuration considered here, a BP layer is placed above a conducting plate, and the distance between them is denoted by $g$ (see Figure 1a). We consider surface plasmons propagating in the $x$ direction. In order to study the influence of BP anisotropy on plasmon propagation, one of the two principal lattice axes is aligned along the $x$ direction ('armchair' ('AC') and 'zigzag' ('ZZ') as illustrated in Figure 1a). Figure 1b shows typical electric field distributions of conventional BP surface plasmons propagating along the AC direction for the free-space wavelength $\lambda_0 = 25$ μm for comparison. Here, we used five layers of BP (thickness $t = 2.675$ nm) and assumed a damping constant of $\eta = 10$ meV and an electron density of $n = 1.0 \times 10^{13}$ cm$^{-2}$. From now on, we will use the same condition for BP and the geometry unless mentioned otherwise. The conductivities we used for numerical simulations are summarized in the supplementary material. Throughout the paper, we will mostly focus on the case of five layers due to their experimental feasibility and reproducibility.[43] However, we will still visit the cases of different numbers of layer for completeness. Conventional BP plasmons exhibit a symmetric field profile with the plasmon wavelength $\lambda_c = 1200$ nm, which gives the vertical confinement of $\lambda_c/2\pi = 191$ nm. Figures 1c and 1d show the field distributions in the presence of a conducting plate for $g = 5$ nm and $\lambda_0 = 25$ μm. As in the case of graphene,[30-32] the electric field is constant across the gap region due to the out-of-phase charge oscillation between the BP layer and the conducting plate, which clearly shows that the observed mode is an acoustic plasmon (the inset in Figure 1d). For the given $\lambda_0$, the vertical confinement of the acoustic plasmon in the positive $z$ direction is $\lambda_0/5,000$, which is around 38 times higher as compared to conventional plasmons propagating in the AC direction and 3 times higher than that of plasmons propagating in the ZZ direction from our numerical results. Contrary to the graphene case, however, the acoustic plasmon wavelength $\lambda_{ac}$ largely differs for the two orthogonal directions, which implicates a strong in-plane anisotropy in the plasmon dispersion for the BP case.

Before numerical investigation, we derive an analytical expression for the plasmon dispersion. The details for the mathematical derivation can be found in the supplementary material. Here, we define the dimensionless momentum $q$ as $k/k_0$ with the plasmon wavenumber in the $x$ direction $k$ and the free-space wavenumber $k_0 \equiv \omega/c$, with $c$ being the speed of light in free-space. Thus, Re($q$) directly gives the ratio of $\lambda_0/\lambda_{ac}$. For Re($q$)$\hat{g} \ll 1$, the plasmon dispersion for a BP layer on a freestanding plane in a vacuum is given as follows.

$$q = \frac{i}{4\alpha} + \sqrt{\left(\frac{i}{4\alpha}\right)^2 + \frac{i}{2\alpha\hat{g}}}, \quad (1)$$

with the dimensionless conductivity $\alpha \equiv (2\pi\sigma)/c$ and the dimensionless gap height $\hat{g} \equiv k_0 g$. The in-plane anisotropy of BP is accounted for by using the anisotropic conductivity $\sigma$ (in Gaussian units).[34]

$$\sigma = \sigma_{AC}\cos^2\theta + \sigma_{ZZ}\sin^2\theta, \quad (2)$$

Here $\theta$ is the angle of propagation direction with the AC axis. We show that at the low frequencies satisfying $\omega \ll 4\sqrt{D/g}$ with the anisotropic Drude weight $D = D_{AC}\cos^2\theta + D_{ZZ}\sin^2\theta$, the plasmon dispersion in Eq. (1) is further simplified, which is

$$q = \frac{1}{\sqrt{-2i\alpha\hat{g}}}. \quad (3)$$

Under the additional assumption of $\eta/\hbar \ll \omega$, Eq. (3) becomes $c/\sqrt{4gD}$ so that $q$ becomes constant in $\omega$ and scales with $g$ as $g^{-1/2}$. Thus, the plasmon dispersion in Eq. (3) clearly shows the characteristic features of acoustic plasmons. Note that constant $q$ in $\omega$ corresponds to linear scaling with $\omega$ in terms of $k$ since $k = q\omega/c$. The linear scaling regime is accordingly given by the intersection of three inequalities; (1) $\omega < 2\sqrt{D/t}$ ($\equiv \omega_{pl}$), (2) Re($q$)$\hat{g} \ll 1$, (3) $\eta/\hbar \ll \omega \ll 4\sqrt{D/g}$. The inequality (1) comes from the condition for the existence of plasmons, Re($\varepsilon_{BP}$) < 0, with the effective permittivity of BP, $\varepsilon_{BP} = 1+i(4\pi\sigma)/(\omega t)$, where the first (second) term denotes dielectric (Drude) response. In the limit of zero thickness, the former is negligible, and its contribution increases with thickness. Similarly, a reduced doping will also enhance the relative dielectric contribution.[44] Lastly, let us consider the two cases outside of linear scaling regime. At $0 \leq \omega \lesssim \eta/\hbar$, the plasmon dispersion in Eq. (3) scales with $\omega$ roughly as $\omega^{-1/2}$ due to the increase in Re($\sigma$). At high frequencies where $1 \lesssim \text{Re}(q)\hat{g}$, on the other hand, the plasmon dispersion asymptotically approaches that of conventional plasmons without a conducting plate and is given by Eq. (S10) in the limit of small $t$.

Plasmon dispersion curves for five layers of BP from numerical simulations along with those from Eq. (1) in the case of $g$ = 5 nm are shown in Figure 2a. In addition, the plasmon dispersion for the case of conventional plasmons without a conducting plate are plotted together for comparison. As shown in the figure, generally, Re($q$) for the AC direction is smaller at a given $\omega$ because of larger $\sigma$. In the case of conventional plasmons without a conducting plate, Re($q$) for the AC direction follows the classical quadratic scaling behavior (which corresponds to linear scaling with $\omega$ in terms of $q$), while for the ZZ direction, it largely deviates from that. This is attributed to a finite $t$, which limits the plasmon supporting frequency up to $\omega_{pl}$ = (0.175 eV)/$\hbar$. At around $\omega = \omega_{pl}$, Re($q$) rapidly increases. The AC case shows no such tendency since $\omega_{pl}$ corresponds to higher frequency, (0.593 eV)/$\hbar$. In the presence of a conducting plate, the plasmon dispersion for the AC direction is nearly constant in $\omega$ at most of frequencies showing the characteristic scaling behavior of acoustic plasmons except for the near-infrared (IR) regime and very small frequencies satisfying $0 \leq \omega \lesssim \eta/\hbar$. In the near-IR regime, $\lambda_{ac}$ is comparable to $g$ so that the plasmon dispersion follows the conventional case. For $0 \leq \omega \lesssim \eta/\hbar$, Re($q$) increases with decreasing $\omega$ as $\omega^{-1/2}$ as expected from Eq. (3), and this is the same for the ZZ direction. Physically, such scaling behaviors come from the absorption, which in the Drude contribution dominates when $\omega \lesssim \eta/\hbar$. These modes are overdamped oscillations, as the real part of conductivity becomes very large at such low frequencies. This leads to the divergent behavior as observed in Figure 2a. In contrast to the AC case, however, the plasmon dispersion for the ZZ direction shows no linear scaling behavior owing to larger Re($q$) and smaller $\omega_{pl}$. As $\omega \to$ (0.175 eV)/$\hbar$, it follows that for the case without a conducting plate. The larger Re($q$) for the ZZ direction also results in a significant discrepancy between the numerical and the analytical results for the acoustic dispersion as well, while the two results for the AC direction are in a good agreement because of a smaller Re($q$).

Figure 2b shows the figure of merit (FOM), Re($q$)/Im($q$), for the plasmon dispersions given in Figure 2a. As expected from Eq. (S15), the FOM increases almost linearly with $\omega$ at small frequencies and the FOMs for different crystal axes have the same value. At higher frequencies, however, it starts to deviate from this trend before rolling down with increasing $\omega$ as intraband Landau damping sets in. Particularly for the ZZ direction, FOM becomes zero at around $\omega =$ (0.175 eV)/$\hbar$. In contrast, the AC case is found to be less damped and persists up to the near-infrared regime due to lower plasmon confinement and broader plasmon-supporting band $\omega <$ (0.593 eV)/$\hbar$. The numerical results also agree with the analytical results in that the FOM for acoustic plasmons is always larger than those for conventional plasmons at any frequency when Re($q$)$\hat{g} \ll 1$. In Figures 2c and 2d, we examined the effect of $g$ on Re($q$) and FOM given $\omega =$ 0.025 eV. For a small $g$, where Re($q$)$\hat{g} \ll 1$, Re($q$) scales with $g$ as $g^{-1/2}$ since it follows Eq. (3). For a large $g$, the analytical expression deviates from the numerical results and Re($q$) asymptotically approaches the conventional case. Figure 2d shows the decrease in the FOM as the plasmon nature changes from acoustic to conventional.

From a practical viewpoint, it is important to consider the effect of the electron density, $n$, as well as the number of layers, $N$, on the acoustic plasmon dispersion, since many potential applications require an active tuning of the optical properties of 2D materials. The plasmon dispersion at different $n$ of $0.5\times10^{13}$, $1.0\times10^{13}$, and $2.0\times10^{13}$ cm$^{-2}$ is shown in Figure 3a. With increasing $n$, Re($q$) decreases at a fixed $\omega$ due to the increase in $D$ and accordingly $\sigma$.[34] In addition, the increase in $D$ broadens the plasmon-supporting band limited by $\omega_{pl}$. In the ZZ case, in particular, $\omega_{pl}$ is located within the frequency regime of interest leading to substantial change in the dispersion with $n$. In the AC case, $\omega_{pl}$ is in the near-IR regime so that the plasmon dispersion is less sensitive to $n$. For both directions, the FOM increases with increasing $n$ at a given $\omega$ because of lower plasmon confinement (Figure 3b).

Shown in Figure 3c is how the acoustic plasmon dispersion changes with increasing $N$ for $g$ = 5 nm. Here, we fixed $n$ to be $1.0\times10^{13}$ cm$^{-2}$. The increase in $N$ accompanies the increase in $D$ owing to the decrease in the in-plane effective mass.[34] For the AC direction, such tendency is manifested through a slight decrease in Re($q$). The increase of Re($q$) for $N$ = 20 in the near-IR regime is attributed to the decrease in $\omega_{pl}$. The change in $\omega_{pl}$ leads to the significant changes in the plasmon dispersion for the ZZ direction as well. With decreasing $N$, the FOMs have a larger value for a broader frequency range (Figure 3d). Note that the results in Figure 3 indicate the larger asymmetry in the dispersion for small $n$ and larger $N$, which agrees well with the expectation from the anisotropy in $\sigma$ (Figure S1c).

In addition to the plasmon dispersion, the modal reflection of plasmons plays an important role in predicting the behavior of plasmonic resonators.[45] In this regard, we investigated the reflectance and reflection phase picked up by acoustic plasmons at the open edge of a BP/free-space/conducting plate system. Here, we focus on two different types of edge termination; semi-infinite BP/semi-infinite conducting plate (SBSC) and infinite BP sheet/semi-infinite conducting plate (IBSC), which are schematically illustrated in Figures 4a and 4b. We exclude the case of a continuous conducting plate due to the inefficient coupling to plasmons under plane wave excitation. Figures 4c and 4d show the electric field distributions along the AC direction after reflection at the SBSC and IBSC edges for $g = 5$ nm and $\lambda_0 = 25$ μm case. In the SBSC case, acoustic plasmons are almost totally reflected, while in the IBSC case, they are coupled to conventional BP plasmons in the metal-free region.

Figures 4e and 4f show the reflectance and reflection phase picked up by plasmons after reflection at the SBSC and IBSC edges for different $g$ of 2, 5, and 10 nm. In the SBSC case, the reflectance of acoustic plasmons is always close to unity due to the absence of a waveguide mode in free-space, similar to the graphene ribbon case.[45] For the ZZ direction where $\lambda_{ac}$ becomes comparable to $t$, the reflectance is smaller due to more efficient coupling to photonic radiation modes at the edge. In the long wavelength limit where acoustic plasmons are more confined within the gap, the reflection phase approaches that for the metal gap plasmon case, $-\pi$, due to the similarity in the mode profile.[46, 47] With increasing $\omega$, the reflection phase converges to a value around $-0.5\pi$. For the ZZ direction, the reflection phase converges to this value more rapidly as $\omega \to (0.175 \text{ eV})/\hbar$. Note that the non-trivial reflection phase obtained here is somewhat different from that for the monolayer graphene case, $-3/4\pi$,[45] because of the larger thickness of five layers of BP. Our numerical results show that $-3/4\pi$ is recovered for a monolayer of BP, indicating the importance of mode profile on the reflection phase.

In contrast to the SBSC case, the reflectance for the IBSC case abruptly decreases with increasing $\omega$ so that at high frequencies, the reflection becomes negligible (Figure 4f). This is because of the small difference in $q$ between acoustic and conventional plasmons, which also accounts for the smaller reflectance for the cases of propagation in the ZZ direction and larger gaps. The reflection phase scales with $\omega$ similarly to the SBSC case, except that it has larger values at most frequencies and converges more rapidly (Figure 6f). Note that we neglect the reflection phases for the ZZ direction at around $\omega = \omega_{pl}$, since the numerical simulations fails to give reliable values for the reflection phase because of a negligibly small $\lambda_{ac}$. This tendency is more severe for IBSC case owing to negligible reflectance.

From our numerical results for the plasmon dispersion and the reflection phase, we estimate the resonant frequencies of two different types of acoustic plasmon resonators, having either periodic ribbons or a continuous BP sheet on a periodic array of conducting plates (gold). These two configurations are considered due to their experimental feasibility. The other feasible design, BP ribbons on a continuous conducting plate, is excluded due to its small far-field signal compared to the other configurations (Figure S4). We also focus only on the AC case to compare the resonant behaviors of different configurations. The numerical results for the ZZ direction can be found in the supplementary material. Figure 5a shows the far-field extinction spectra for the case of BP(ribbon)/metal(ribbon) with $g = 5$ nm as a function of inverse conducting plate width ($1/w$). We set the distance between two neighboring gold plates to be $w$ as well. The far-field extinction is defined by $1-T$ with $T$ being normalized far-field transmittance. The magenta dashed lines show the estimated resonant frequencies from plasmon dispersion in Figure 2a and the reflection phase shown in Figure 4e using the Fabry-Perot equation, which is given as

$$2kw + 2\Phi_r = 2m\pi, \qquad (4)$$

where $\Phi_r$ and $m$ represent the reflection phase at the edge and the order of the Fabry-Perot resonance, respectively. As shown in the figure, the estimated resonant frequencies for different Fabry-Perot resonance orders agree very well with full numerical results, indicating weak coupling between two neighboring resonators. Figure 5b shows extinction spectra for the case of BP(sheet)/metal(ribbon). In contrast to the previous case, the estimated resonant frequencies using the reflection phase in Figure 4f significantly deviates from full numerical results. In the continuous BP case, the unit cell of the plasmon resonator should be considered as a combination of two Fabry-Perot resonators formed within the gap and intermediate region. Based on such an assumption, we develop a new Fabry-Perot resonance model to account for the resonant behavior. The details for our model can be found in the supplementary material. According to our new model, the resonance condition is given as

$$2k_{ac}w + 2k_c w = 2l\pi, \qquad (5)$$

Here, $k_{ac}$ and $k_c$ represent the wavenumber of acoustic and conventional plasmons, respectively, and $l$ is the order of Fabry-Perot resonances. As shown in the figure, the resonance frequency estimated from the new model give a decent match with numerical results. In contrast to a conventional model, a zeroth-order Fabry-Perot resonance is not allowed due to the absence of the phase term in Eq. (5) and accordingly the first occurring mode corresponds to the second-order Fabry-Perot resonance. As a result, the plasmon resonances can occur at higher frequencies than the case of BP(ribbon)/metal(ribbon).

Based on the resonance model developed above, we investigate the extinction intensities of two different acoustic plasmon resonators together with a conventional BP ribbon resonator. Figures 6a-c illustrate optical coupling routes in three different designs. In the BP(ribbon) and BP(ribbon)/metal(ribbon) cases, incident waves are coupled to conventional surface plasmons or acoustic plasmons directly after scattering at the edge of the resonators. In the BP(sheet)/metal(ribbon) case, however, the coupling of free-space light to acoustic plasmons can occur not only in a direct manner but also after being coupled to conventional plasmons. The electric field enhancement maps on the first occurring resonances ($\omega$ = 0.083 eV/$\hbar$) for three different designs are shown in Figures 6d-f. Figures 6e and 6f clearly show that the resonances indeed result from acoustic plasmons. The largest enhancement for the BP(sheet)/metal(ribbon) case indicates the stronger coupling compared to the other configurations. Figure 6g shows the extinction spectra for three different configurations where the first-occurring resonances coincide at $\omega = 0.083$ eV/$\hbar$. The smaller extinction for the BP(ribbon)/metal(ribbon) case than the ribbon case implicates the low efficiency of direct coupling to acoustic plasmons compared to that to conventional plasmons, which results from the larger difference in wavenumber with plane wave. The BP(sheet)/metal(ribbon) case indeed exhibits the largest extinction as expected. In contrast to the ribbon case, high-order modes also have substantial extinction intensities on resonance due to the smaller increase in $w$ and narrower spectral distances between different orders. We found out that the trend in Figure 6g between three designs holds at mid-IR and longer wavelength regimes, which cover most of frequency range of interest (Figure 6h). With the same BP material parameters, the BP(sheet)/metal(ribbon) case exhibits the largest extinction on the first-occurring resonance of up to 60% for $\omega < 0.142$ eV/$\hbar$. Using the plasmon dispersion and the extinction results in Figure 6g, we extract the coupling efficiency $\kappa$ for different designs as shown in the inset (The details for calculation can be found in the supplementary material.). The inset clearly shows that the larger extinction for the BP(ribbon)/metal(ribbon) case is mainly attributed to the higher coupling efficiency. Our numerical results show that, among the two main contributors to the extinction intensity, the quality of cavity (including plasmon propagation loss and reflectance at resonator edges) and coupling efficiency, coupling efficiency plays a more important role. The increase in $\kappa$ with decreasing $\omega$ for the BP ribbon case is attributed to the decrease in Re($q$) while the opposite scaling behavior for the BP(ribbon)/metal(ribbon) case comes from the increase in Re($q$) of acoustic plasmons. The similarity in the scaling behavior between the BP ribbon and BP(sheet)/metal(ribbon) cases implies the dominant role of the indirect coupling from conventional plasmons to acoustic plasmons. In addition to higher extinction intensity, the BP(sheet)/metal(ribbon) resonators are more suitable for practical implementation compared to the ribbon and the BP(ribbon)/metal(ribbon) type resonators since potentially, no patterning steps are required after the deposition of BP, thereby minimizing unwanted process-induced damages to BP.

In conclusion, we have investigated the anisotropic dispersion for the acoustic plasmons in a monolayer and multilayers of freestanding BP coupled to a conducting plate. The dispersion for the acoustic plasmons was found to scale linearly with $\omega$ in the mid- and far-IR regime except in the long wavelength limit. At high frequencies, where $\lambda_{ac}$ becomes comparable to $g$, it approaches the dispersion of conventional BP plasmons without a conducting plate. Due to larger confinement and narrower plasmon-supporting band, the ZZ case largely deviates from the linear scaling behavior. The analytical results confirmed the numerical results and clearly showed that the linear scaling regime become broader for smaller gap size and number of layers, and higher carrier density. Further, we numerically demonstrated different types of acoustic plasmon resonators including BP(ribbon)/metal(ribbon) and BP(sheet)/metal(ribbon) configurations. Among feasible design options, the BP(sheet)/metal(ribbon) resonator exhibited the largest extinction intensity than the other possible configurations due to higher coupling efficiency. We developed a modified Fabry-Perot resonance model to account for the resonant behavior of such a plasmon resonator. Importantly, our new resonator design can be realized using a continuous sheet of BP without nano-patterning, which can introduce defects and edge roughness in BP. While an experimental realization of acoustic plasmon resonances in BP is not trivial, recent advances[43, 48] in the growth of large-area high-quality BP samples show promising routes for the verification of our theoretical predictions. Also, our findings on acoustic plasmons in BP help to develop novel optoelectronic devices using optical anisotropy such as metasurfaces, photodetectors, high-resolution imaging systems, and biosensors.

**METHODS**

We used COMSOL Multiphysics with the RF Module for numerical simulations. In order to calculate the dispersion relation for acoustic plasmons in BP along with the reflection phase and amplitude, a port placed inside the simulation domain was used to solve for the eigenmode, launch the mode, and measure the reflection from the terminal interface. Perfectly matched layers (PMLs) were used at all simulation boundaries to increase accuracy. The electric field distributions in Figure 1 and Figure 4 were also calculated in the same configuration. For Figure 5 and Figure 6, we used a plane wave with transverse magnetic polarization to obtain the extinction spectra of two acoustic plasmon resonators. Perfect electrical conductor (PEC) boundary conditions were used at both boundaries to simulate a periodic structure and reduce the computation time through symmetry. In all cases, the conducting plate was assumed to be 50 nm-thick gold with a dielectric function obtained elsewhere.[49] Except for Figure 3, five layers of BP with $n = 1.0\times10^{13}$ cm$^{-2}$, $T = 300$ K, and $\eta = 10$ meV was considered and its conductivities were calculated using the Drude equation in Eq. (S1) with $D_x = 5.438\times10^{20}$ and $D_y = 4.769\times10^{19}$ m·s$^{-2}$ (in Gaussian units). The details for the calculation of the Drude weight can be found in the supplementary material. The BP layer was modeled as a 2.675 nm-thick slab with the effective permittivity obtained from its conductivity.[44]

**AUTHOR CONTRIBUTIONS.**

I.-H.L., L.M.M., and S.-H.O. conceived the idea. I.-H.L. and D.A.M. performed numerical simulations using COMSOL. I.-H.L., L.M.M., K.K., and T.L contributed to theoretical analysis. K.K. and T.L. calculated material parameters. All authors analyzed the data and wrote the paper together.

**ACKNOWLEDGMENTS.**

This research was supported primarily by the U.S. National Science Foundation (MRSEC Seed grant to I.-H.L., T.L., K.K., S.-H.O.; ECCS 1610333 to S.-H.O.). L.M.-M. acknowledges financial support by the Spanish MINECO under contract No. MAT2014-53432-C5. D.A.M. acknowledges the National Institutes of Health Biotechnology Training Grant (NIH T32 GM008347). L.M.-M., T.L., and S.-H.O. also thank support from the Institute for Mathematics and its Applications (IMA) at the University of Minnesota.

**Notes.** The authors declare no competing financial interests.

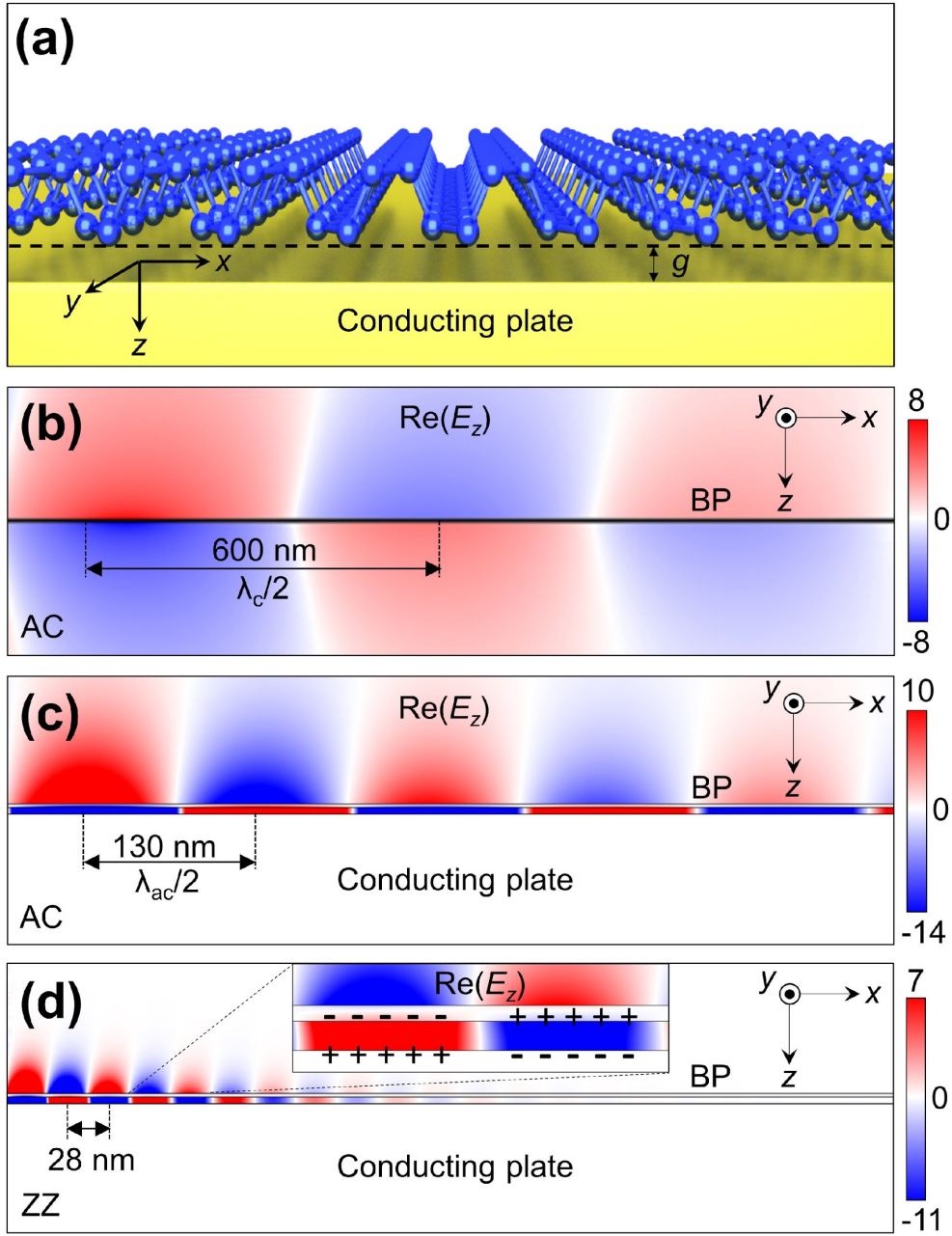

**Figure 1.** (a) The geometrical configuration supporting acoustic plasmons. The $z$-component of the electric field of (b) conventional plasmons propagating in the AC direction with the plasmon wavelength, $\lambda_c$, and acoustic plasmons propagating in (c) the AC and (d) ZZ directions are shown. The inset in (d) shows the magnified image of the electric field for one plasmon wavelength, $\lambda_{ac}$, to illustrate how charges oscillate. Here, the gap between the BP and the conducting plate $g$ was 5 nm, and the free-space wavelength was $\lambda_0 = 25$ μm.

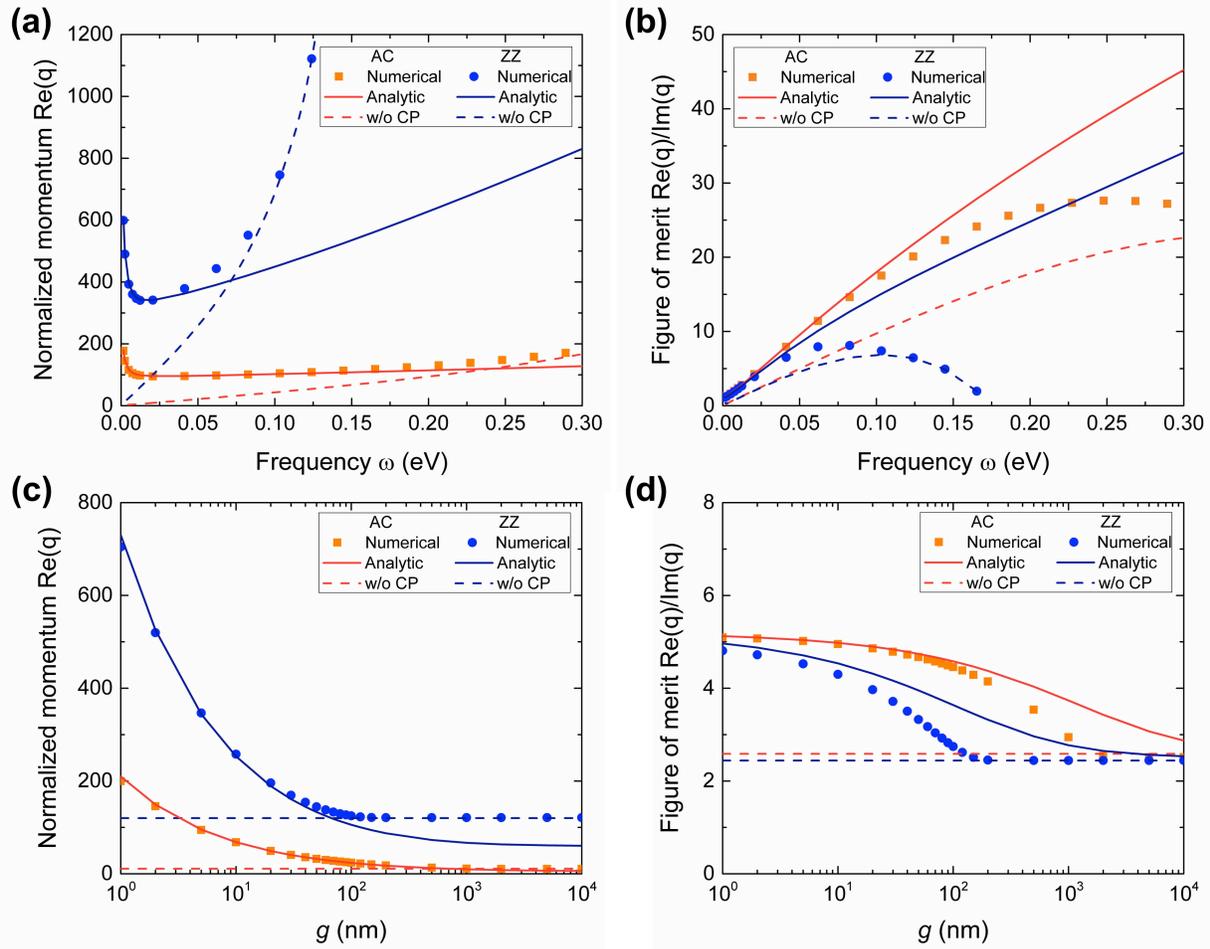

**Figure 2.** (a) The plasmon dispersion and (b) figure of merit (FOM) along the AC and ZZ directions. Here, we assumed $g = 5$ nm. (c) The plasmon dispersion and (d) FOM as a function of $g$ along the AC and ZZ directions. In all cases, the dispersions given in terms of the real part of the dimensionless momentum, $q \equiv k/k_0$ and FOM is defined as $Re(q)/Im(q)$. Numerical and analytical results are plotted together, and the numerical results for plasmons without a conducting plate (CP) are given for comparison.

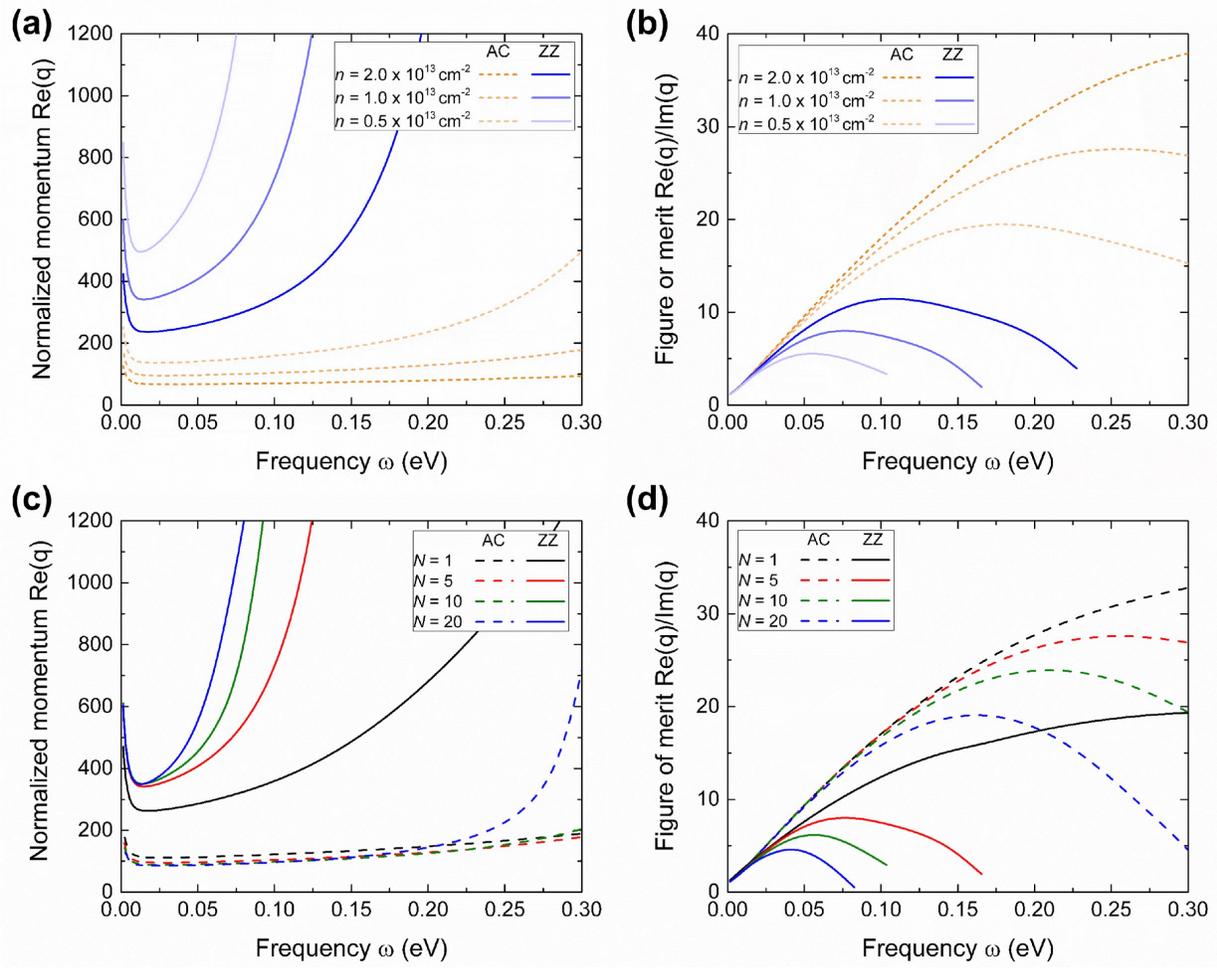

**Figure 3.** Effect of doping on (a) plasmon dispersion and (b) FOM. Electron density is varied from $0.5 \times 10^{13}$, $1.0 \times 10^{13}$ to $2.0 \times 10^{13}$ cm$^{-2}$. The effect of the number of layers on (c) plasmon dispersion and (d) FOM. Here, 1, 5, 10, and 20 layers are considered.

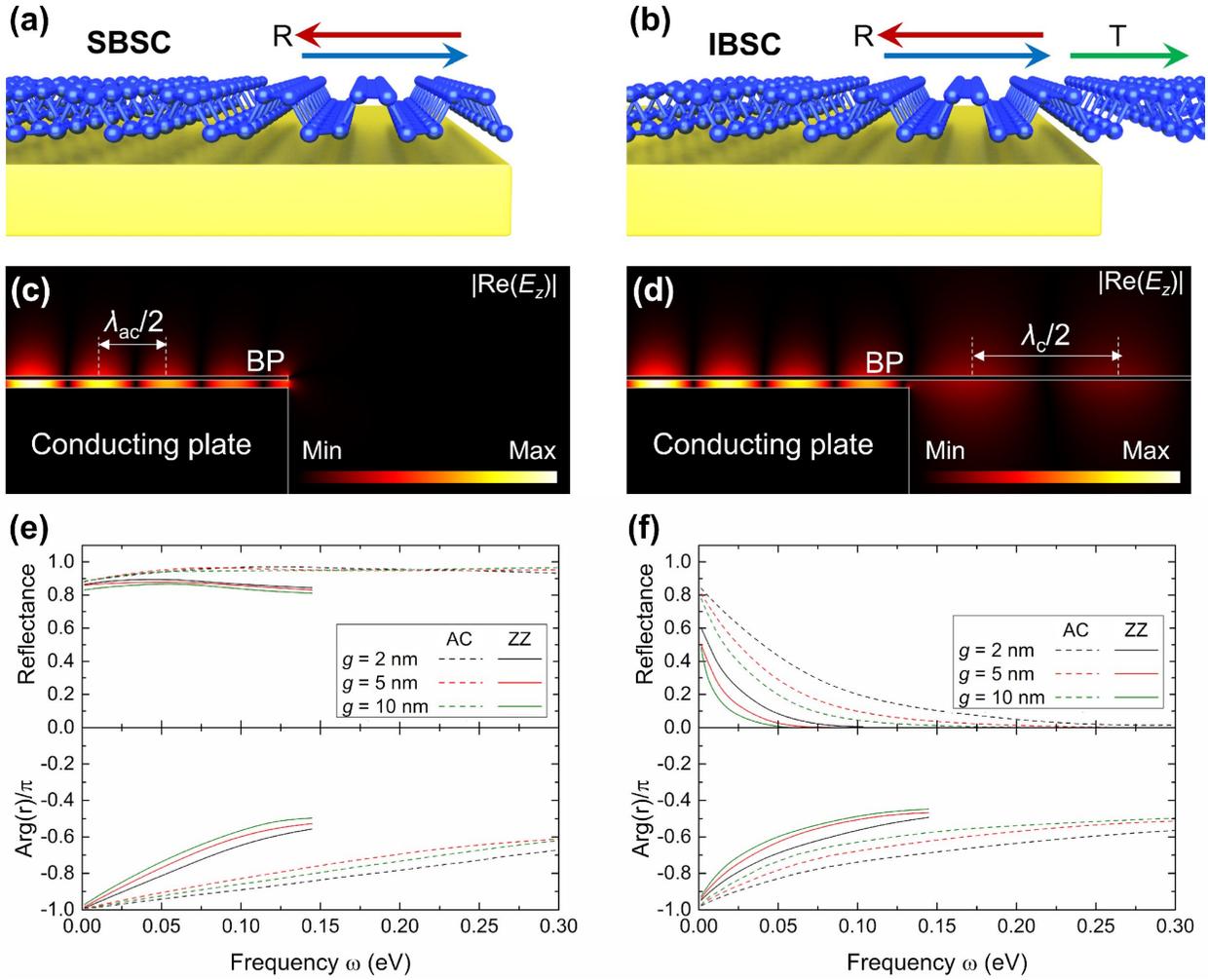

**Figure 4.** Schematic illustration of reflection at two different types of edge termination with (a) semi-infinite and (b) infinite BP sheet over the edge of a semi-infinite conducting plate (SBSC and IBSC cases, respectively). Electrical field distribution after reflection for (c) SBSC and (d) IBSC cases for AC/g = 5 nm/$\lambda_0$ = 25 μm. Reflection amplitude and phase of an acoustic plasmon after reflection at the edge for (e) SBSC and (f) IBSC cases for different $g$ and crystal axes.

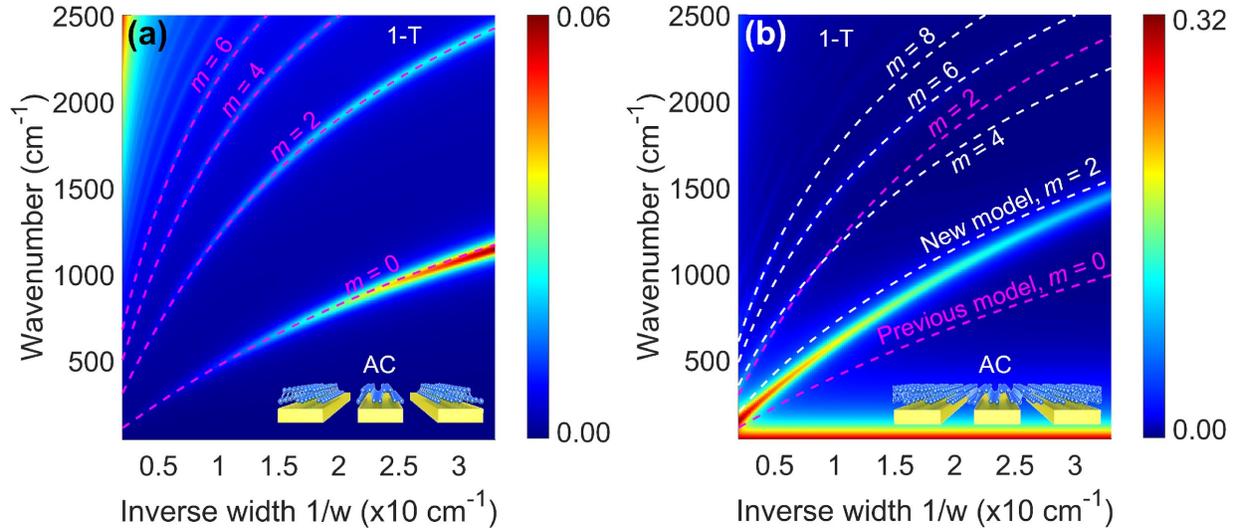

**Figure 5.** Extinction spectra of BP acoustic plasmon resonators with (a) BP(ribbon)/metal(ribbon) and BP(sheet)/metal(ribbon) for the AC direction as a function of inverse conducting plate width ($1/w$) under illumination by a normally incident plane wave with transverse magnetic polarization. Magenta dashed lines are the estimated resonant frequencies for different orders of interference ($m$) using the conventional Fabry-Perot resonance equation from the plasmon dispersion and reflection phase obtained previously. White dashed lines are the estimation from the modified Fabry-Perot resonance equation.

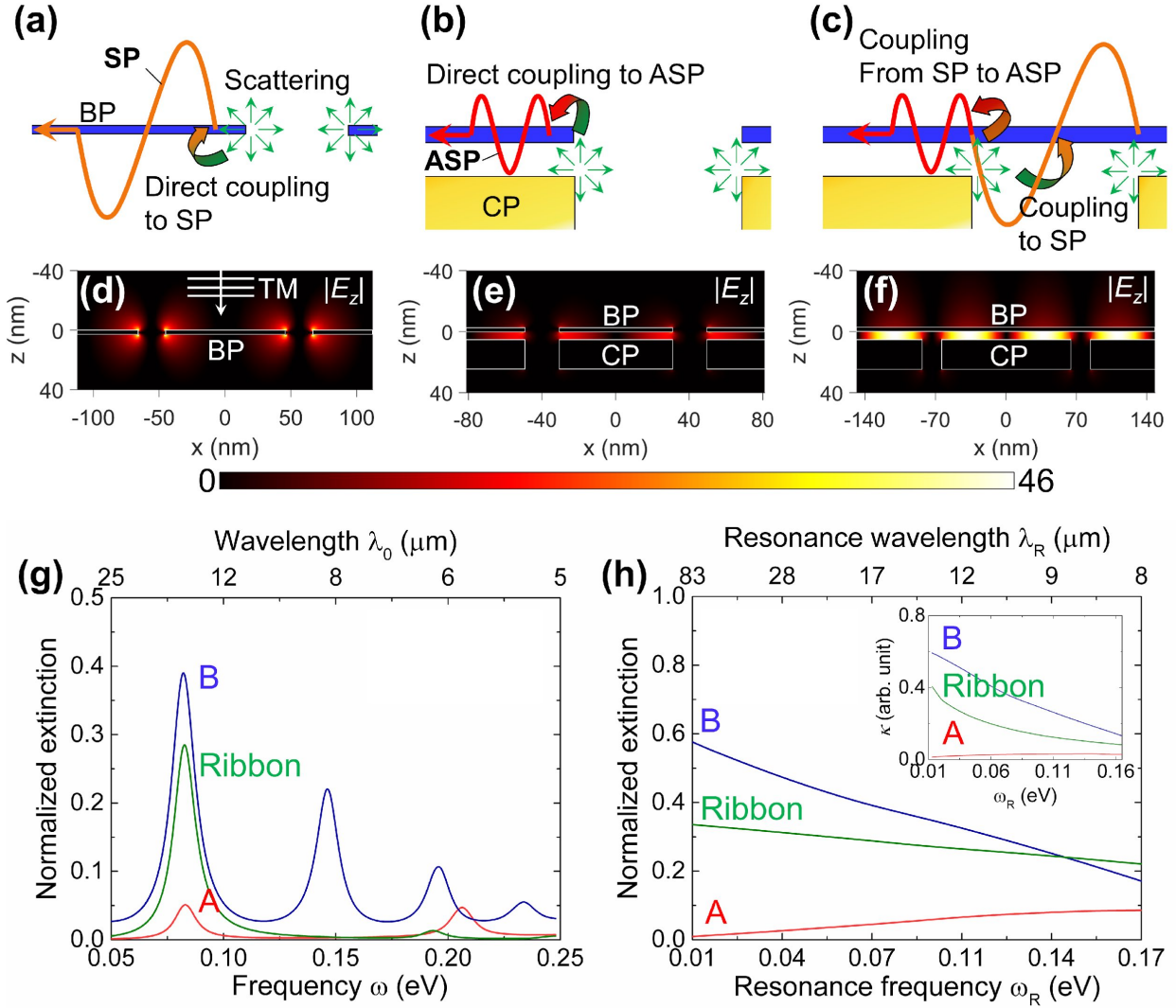

**Figure 6.** Schematic illustration of coupling routes to surface plasmons for (a) BP(ribbon), (b) BP(ribbon)/metal(ribbon) ('A'), and BP(sheet)/metal(ribbon) ('B') resonators. The $z$ electric field enhancement on resonance ($\lambda_0 = 15$ μm) for (a) BP(ribbon), (b) A, and B resonators. (g) Extinction spectra for three different geometries where the first-occurring resonance coincide at $\lambda_0 = 15$ μm. In (d)-(g), the widths of a resonator unit were 92, 61, and 128 nm, respectively. (h) Normalized extinction intensity on resonance as a function of resonance frequency for three different geometries. Inset shows corresponding coupling efficiency $\kappa$. In (d)-(h), the spacing between two resonator units was fixed to be 20 nm and the thickness of a conducting plate was 20 nm. The resonators were illuminated by a normally incident plane wave with transverse magnetic (TM) polarization.

SUPPLEMENTRAY INFORMATION

# Anisotropic Acoustic Plasmons in Black Phosphorus

In-Ho Lee[1], Luis Martin-Moreno[2,]*, Daniel A. Mohr[1], Kaveh Khaliji[1],

Tony Low[1,]*, and Sang-Hyun Oh[1,]*

[1] Department of Electrical and Computer Engineering, University of Minnesota, Minneapolis, Minnesota, 55455, U.S.A.

[2] Instituto de Ciencia de Materiales de Aragón and Departamento de Física de la Materia Condensada, CSIC-Universidad de Zaragoza, E-50009 Zaragoza, Spain.

*E-mail: lmm@unizar.es (L.M.-M.), tlow@umn.edu (T.L.), sang@umn.edu (S.-H.O.)

# 1. Dynamic conductivity of multilayer BP

All simulations were carried out for photon energies of up to 0.3 eV, the bandgap of bulk BP. Within this energy window, the Drude model would suffice to model the conductivity of multilayer BP system,[1]

$$\sigma_j = \frac{iD_j}{\omega + i\eta/\hbar}, \qquad D_j = e^2 \sum_i \frac{n_i}{m_i^j} \qquad (S1)$$

where $D_j$ is the Drude weight along the $j$-axis with $j$ = AC or ZZ. $m_i^j$ is the effective mass of the $i$-th conduction subband along the $j$-axis,[1, 2]

$$m_i^x = \frac{\hbar^2}{2(\eta_i^x + \gamma_i^x + \chi_i^2/(2\delta_i))}, \qquad m_i^y = \frac{\hbar^2}{2(\eta_i^y + \gamma_i^y)} \qquad (S2)$$

where $\eta_i^x$ (in units of eV·Å$^2$) = 0.364$\Lambda_i$ − 1.384, $\gamma_i^x$ (eV·Å$^2$) = 2.443 $\Lambda_i$ + 2.035, $\chi_i$ (eV·Å) = 2.071 $\Lambda_i$ + 5.896, $\delta_i$ (eV) = 0.712 $\Lambda_i$ + 0.919, and $\Lambda_i$ = cos[$i\pi/(N+1)$], with $N$ being the number of layers. The electron density of the $i$-th conduction subband, $n_i$, is given by,[3]

$$n_i = \frac{m_i^{dos} k_B T}{\pi\hbar} \log\left[1 + \exp\left(\frac{\mu - E_i}{k_B T}\right)\right], \qquad (S3)$$

where $m_i^{dos} = \sqrt{m_i^x m_i^y}$ is the density of states effective mass. $E_i$ (eV) = 0.58$\Lambda_i$ + 0.505 is the minimum energy of the $i$-th conduction subband. $\mu$ is the chemical potential, reported relative to the energy of the first conduction subband. The total electron density is given by $n = \sum_i n_i$.

In Figure S1a, the dynamic conductivity used in the calculations of Figures 3a and 3b of the main text, are shown as an example. The dynamic anisotropy is quantified via the ratio, $D_x/D_y$. The variation of this ratio with number of layers and electron density are recorded in Figure S1b. It is apparent that the BP anisotropy is maximum for thicker samples with low doping concentration.

Our calculations indicate that the armchair-to-zigzag Drude weight ratio can increase to more than 14, for 10-layer thick BP and $10^{12}$ cm$^{-2}$ electron density.

## 2. Acoustic plasmon dispersion

Here we will derive an analytic expression for acoustic plasmon dispersion assuming that (1) the thickness of 2D materials is negligible and (2) the conducting plate is perfect electrical conductor. Before beginning, it should be noted that the following discussion does not assume a specific 2D material so that analytic expression obtained here is general for all 2D plasmonic materials. As illustrated in Figure S2, a 2D material layer is placed at $z = 0$ and free-standing in a vacuum with the dielectric constant of a unity. The top of conducting plate corresponds to $z = g$. In this case, the electromagnetic fields in two regions can be represented as

$$\begin{cases} E_x(x,z) = e^{ikx}e^{ik_z z} + r_k e^{ikx}e^{-ik_z z}, \\ -\vec{u}_z \times \vec{H}(x,z) = Y e^{ikx}e^{ik_z z} - Y r_k e^{ikx}e^{-ik_z z}, \end{cases} \quad -\infty < z < 0$$

$$\begin{cases} E_x(x,z) = A e^{ikx}e^{ik_z z} + B e^{ikx}e^{-ik_z z}, \\ -\vec{u}_z \times \vec{H}(x,z) = Y A e^{ikx}e^{ik_z z} - Y B e^{ikx}e^{-ik_z z}, \end{cases} \quad 0 < z < g \tag{S4}$$

where $k$ is the wavenumber component in $x$ direction. Here, we define the free-space wavenumber $k_0 \equiv \omega/c$ and the admittance $Y \equiv k_0/k_z$ with the speed of light in free-space $c$ and $k_z = \sqrt{k_0^2 - k^2}$. The continuity of the tangential component of $\vec{H}$ across the 2D system statisfies

$$\vec{u}_z \times \left[\vec{H}(z=0^+) - \vec{H}(z=0^-)\right] = \frac{4\pi}{c} \vec{j}(z=0) = 2\alpha \vec{E}_\parallel(z=0), \tag{S5}$$

where $\alpha \equiv (2\pi\sigma)/c$ with the conductivity of a 2D material $\sigma$, as given in Eq. (S1). At this point, it is convenient to define the dimesionless momenta; $q = k/k_0$ and $q_z = k_z/k_0$, and dimensionless gap size as $\hat{g} \equiv k_0 g$. Note that Re($q$) directly gives the ratio of $\lambda_0/\lambda_{pl}$ with the free-space wavelength $\lambda_0$ and the plasmon wavelength $\lambda_{pl}$. From the electromagnetic field boundary conditions for two interfaces ($z = 0$ and $g$), thus, we obtain

$$\begin{cases} 1 + r_k = A + B, \\ -(A-B)Y + (1-r_k)Y = 2\alpha(1+r_k), \\ A\hat{e} + B\hat{e}^{-1} = 0, \end{cases} \quad (S6)$$

with $\hat{e} = exp(iq_z\hat{g})$. Since $B = -\hat{e}^2 A$ from Eq. (S6), the case without a conducting plate at $z = g$ can be readily considered by taking $\hat{e} = 0$. From three boundary conditions, the dispersion relation is derived as follows.

$$r_k = \frac{(1-\phi)Y - 2\alpha}{(1+\phi)Y + 2\alpha}, \quad (S7)$$

where $\phi \equiv (1+\hat{e}^2)/(1-\hat{e}^2)$. The eigenmode (plasmon) of the system is obtained when the denominator in Eq. (S7) vanishes, which gives

$$(1+\phi)Y + 2\alpha = 0. \quad (S8)$$

Before moving on to the case with a conducting plate, let us consider the simpler case without a conducting plate. In this case, $\hat{e} = 0$ and accordingly, $\phi = 1$. Therefore, we obtain the plasmon dispersion of $q_z = -1/\alpha$. This dispersion relation coincides with that for the conventional 2D plasmons.[4]

Now let us consider the case with a conducting plate. For the 2D plasmons, $k$ is much larger than $k_0$ so that $q_z \approx iq$. Thus, Eq. (S8) becomes

$$\left(1 + \frac{1+e^{-2q\hat{g}}}{1-e^{-2q\hat{g}}}\right)\frac{1}{q} = -2i\alpha. \quad (S9)$$

From Eq. (S9), two limit cases can be readily obtained.

(1) When $1 \ll \text{Re}(q)\hat{g}$,

$exp(-2q\hat{g})$ is close to 0 so the plasmon dispersion is given as

$$q = \frac{i}{\alpha}. \tag{S10}$$

Thus, the dispersion for the conventional plasmons without a conducting plate in the limit of $k_0 \ll k$ is recovered.

(2) When $\text{Re}(q)\hat{g} \ll 1$,

$(1+ exp(-2q\hat{g}))/(1- exp(-2q\hat{g}))$ can be approximated by $1/(q\hat{g})$. Under this assumption, Eq. (S9) becomes the quadratic equation for $q$. By solving the equation for $q$, we obtain the following dispersion relation.

$$q = \frac{i}{4\alpha} + \sqrt{\left(\frac{i}{4\alpha}\right)^2 + \frac{i}{2\alpha\hat{g}}}. \tag{S12}$$

It is worthwhile to visit the case of small frequencies $\omega \ll 4\sqrt{D/g}$. Our numerical results for BP show that this frequency band is usually located across the mid- and far-infrared regime, and for the armchair case, it even persists up to the near-infrared regime. At these frequencies, the right term inside the square root, $i/(2\alpha\hat{g})$, is much larger than the other term, $(i/4\alpha)^2$, in terms of magnitude. Hence, the plasmon dispersion can be approximated to simpler form as follows.

$$q = \frac{1}{\sqrt{-2i\alpha\hat{g}}}. \tag{S13}$$

Let us additionally assume $\eta/\hbar \ll \omega$ so that $\text{Re}(\sigma) \ll \text{Im}(\sigma)$. In this case, $q$ becomes $c/\sqrt{4gD}$, which is constant in $\omega$ and scales with $g$ as $g^{-1/2}$. Note that for $\omega \lesssim \eta/\hbar$, the plasmon dispersion in Eq. (S13) does not linearly scales with $\omega$ but as $\omega^{-1/2}$ since $\text{Re}(\sigma)$ becomes comparable to $\text{Im}(\sigma)$. For anisotropic materials including BP, $\sigma$ will have different values according to the propagation direction of acoustic plasmons and thus, plasmon dispersion becomes anisotropic as well. The in-plane anisotropy can be easily incorporated into the dispersion by simply taking $\sigma$ as

$$\sigma = \sigma_{AC}\cos^2\theta + \sigma_{ZZ}\sin^2\theta, \tag{S14}$$

where $\theta$ is the angle of the in-plane wavevector $\mathbf{k}$ with the armchair axis.

In the light of above discussions, we can derive conditions for the acoustic plasmons to exhibit the linear scaling with $\omega$. The condition for the linear scaling is given by the intersection of the frequency bands given by three inequalities; (1) $\omega < 2\sqrt{D/t}$, (2) $\mathrm{Re}(q)\hat{g} \ll 1$, (3) $\eta/\hbar \ll \omega \ll 4\sqrt{D/g}$. In the more realstic case of BP having finite thickness, the condition for the plasmon existence should be considered as well since the plasmons can be supported only at the frequencies satifying $\mathrm{Re}(\varepsilon_{BP}) < 0$ with the effective permittivity of BP, $\varepsilon_{BP}$. From the relation of $\varepsilon_{BP} = 1+(i4\pi\sigma)/(\omega t)$ and Eq. (S1), it gives the inequality (1), $\omega < 2\sqrt{D/t-(\eta/\hbar)^2} \approx 2\sqrt{D/t}$ ($\equiv \omega_{pl}$).

Before the end of discussion, it would be meaningful to compare the figure-of-merit (FOM) for conventional and acoustic plasmons. From Eq. (S10), the FOM for the conventional case is simply given as $\hbar\omega/\eta$. For the acoustic case, the FOM can be derived from Eq. (S13) as

$$\frac{\mathrm{Re}(q)}{\mathrm{Im}(q)} = \sqrt{\left(\frac{\hbar\omega}{\eta}\right)^2 + 1} + \frac{\hbar\omega}{\eta}. \tag{S15}$$

From Eq. (S15), it is evident that the FOM for acoustic plasmons is always larger than that of conventional plasmons when $\mathrm{Re}(q)\hat{g} \ll 1$. When $\eta/\hbar \ll \omega$, the FOM of acoustic plasmons is appoximated to $2\hbar\omega/\eta$, which is twice the conventional case. Interestingly, the FOM is insensitive to the propagation direction.

## 3. Modified Fabry-Perot model

Let us consider an acoustic plasmon resonator having a continous BP sheet over a periodic array of conducting plates as illustrated in Figure S3a. Upon plane wave incidence, not only acoustic plasmons within Region 1, but also conventional plasmons for the case without a conducting plate in Region 2 are excited. This type of plasmon resonantor can be considered to have two Fabry-Perot resonators in parallel within the unit cell. Therefore, the Fabry-Perot resonace for the whole resonator is obtained when the following two resonance conditions are met at the same frequency.

$$\begin{cases} 2k_{ac}w_1 + 2\Phi_1^* = 2m\pi, \\ 2k_c w_2 + 2\Phi_2^* = 2n\pi, \end{cases} \quad m, n = \text{even integers including } 0 \tag{S16}$$

where $\Phi^*$ represents the phase of $r^*$ which is effective reflection coefficient for plasmons. The subscripts 1 and 2 denote Region 1 and 2. $k_{ac}$ and $k_c$ represent the wavenumber in $x$ direction for acoustic and conventional plasmons, respectively. $r^*$ can be obtained from equivalent perfect electrical conductor models illustrated in Figures S3b and S3c as follows.

$$\begin{cases} r_1^* = r_1 + t_{12}^2 \exp(ik_c w_2) + r_2 t_{12}^2 \exp(i2k_c w_2) + \cdots = r_1 + \dfrac{t_{12}^2 \exp(ik_c w_2)}{1 - r_2 \exp(ik_c w_2)}, \\ r_2^* = r_2 + t_{21}^2 \exp(ik_{ac} w_1) + r_1 t_{21}^2 \exp(i2k_{ac} w_1) + \cdots = r_2 + \dfrac{t_{21}^2 \exp(ik_{ac} w_1)}{1 - r_1 \exp(ik_{ac} w_1)}, \end{cases} \tag{S17}$$

where $r_1$ and $r_2$ represent $r_1^*(w_2 = \infty)$ and $r_2^*(w_1 = \infty)$. In most cases, however, two conditions in Eq. (S16) cannot be satisfied at the same frequency so that no analytic condition can be obtained. In this case, the resonance condition should be obtained numerically from the equation for the total energy stored within the plasmon resonator. Let us instead consider the case when $r_1$ and $r_2$ is very small. This assumption is appropriate at most of frequencies as can be seen in Figure 4f. In this case, $r^*$ is approximated to

$$\begin{cases} r_1^* \approx t_{12}^2 \exp(ik_c w_2), \\ r_2^* \approx t_{21}^2 \exp(ik_{ac} w_1). \end{cases} \tag{S18}$$

Therefore, $\Phi^*$ becomes

$$\begin{cases} \Phi_1^* \approx k_c w_2 + 2\Phi_{12}^t, \\ \Phi_2^* \approx k_{ac} w_1 + 2\Phi_{21}^t. \end{cases} \tag{S19}$$

where $\Phi^t$ denote the phase of $t$. By substituting $\Phi^*$ in Eq. (S16) by Eq. (S19), we get

$$\begin{cases} 2k_{ac}w_1 + 2k_c w_2 + 4\Phi^t_{12} = 2m\pi, \\ 2k_c w_2 + 2k_{ac}w_1 + 4\Phi^t_{21} = 2n\pi, \end{cases} \tag{S20}$$

Since $\Phi^t$ is close to 0, the two conditions in Eq. (S16) can be approximated to

$$2k_{ac}w_1 + 2k_c w_2 = 2l\pi. \qquad l = \text{even integers larger than 0} \tag{S21}$$

In Eq. (S21), the left term corresponds to the round trip of an entire unit cell terminated by PEC at both ends. Since no reflection phase is involved, the minimum order of the resonance is not $l = 0$, but $l = 2$. One of the interesting results from the modified Fabry-Perot resonance condition is that the resonance shifts with $w_2$, which cannot be included in the conventional resonance model. Figure S3d shows the numerical results for the far-field extinction spectrum as a function of $w_2$. The material parameters for BP are the same as in Figure 6a. The numerical results show that the resonant frequencies indeed shift with $w_2$. The resonant frequencies estimated using the modified Fabry-Perot resonance model are in a good agreement with numerical results.

## 4. Extraction of coupling efficiency

Since the BP(ribbon)/metal(ribbon) resonator is well described by the conventional Fabry-Perot resonance model, the extinction intensity on resonance can be modeled as

$$1 - T = \frac{\kappa_A}{1 + |r^*|^4 - 2|r^*|^2 \cdot \exp[-2|\text{Im}(k_{ac})|w_1]}, \tag{S22}$$

where $R = |r^*|^2$ and $\kappa_A$ represents the coupling efficiency from incident wave to acoustic plasmons. In this case, $r^* \approx r^*(w_2 = \infty)$ due to weak coupling between neighboring resonator units. In the BP(sheet)/metal(ribbon) case, Eq. (S22) is modified to

$$1 - T = \frac{\kappa_B}{1 + |t|^8 - 2|t|^4 \cdot \exp[-2|\text{Im}(k_{ac})|w_1 - 2|\text{Im}(k_c)|w_2]}, \tag{S23}$$

where we assume $t = t_{12} = t_{21}$. The coupling efficiecny $\kappa_B$ has two different contributors from acoustic plasmon resonator and conventional plasmon resonator formed between two conducting plate units. Note that in Eqs. (S22) and (S23), the denominators represent the effect from the quaility of cavities including the propagation loss and the reflectance at mirrors. Thus, $\kappa$ represents a contribution only from the coupling efficiency of certain reonator geometry excluding the effect from the quality of cavity. It can be extracted from the numerical results for extinction intensity 1-$T$ using following relations.

$$\begin{cases} \kappa_A = (1-T)\left[1 + |r^*|^4 - 2|r^*|^2 \cdot \exp\{-2|\text{Im}(k_{ac})|w_1\}\right], \\ \kappa_B = (1-T)\left[1 + |t|^8 - 2|t|^4 \cdot \exp\{-2|\text{Im}(k_{ac})|w_1 - 2|\text{Im}(k_c)|w_2\}\right] \end{cases} \quad (S24)$$

Figure S6 shows the contribution from the quality of cavities to extinction intensitiy on resonance for three different resonator designs; BP ribbon, BP(ribbon)/metal(ribbon), and BP(sheet)/metal(ribbon), which is defined as $(1-T)/\kappa$ in Eqs. (S22) and (S23). It shows that the BP(sheet)/metal(ribbon) case has the smallest value, which is totally opposite to the trend of extinction intensity in Figure 6. These results indicate that the coupling efficiency is more important contributor in the extinction intensity than that from cavity quality.

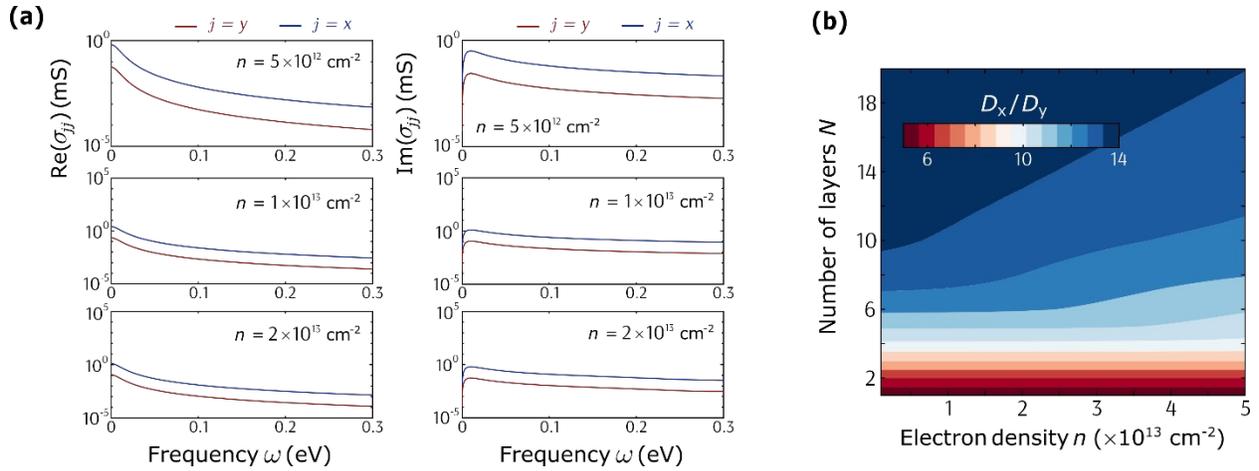

**Figure S1.** (a) The real and (b) imaginary parts of conductivity for five layers of BP. (b) The anisotropy ratio as functions of electron density and number of layers. For all simulations: $T = 300$ K, and $\eta = 10$ meV.

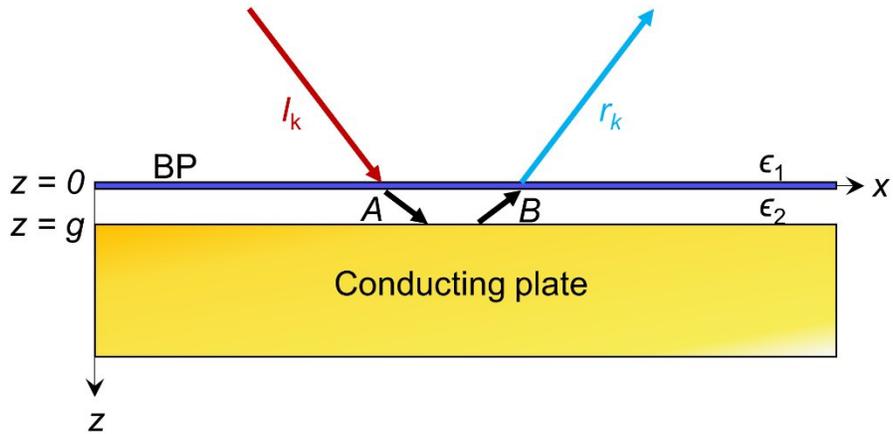

**Figure S2.** Geometry of the electromagnetic problem.

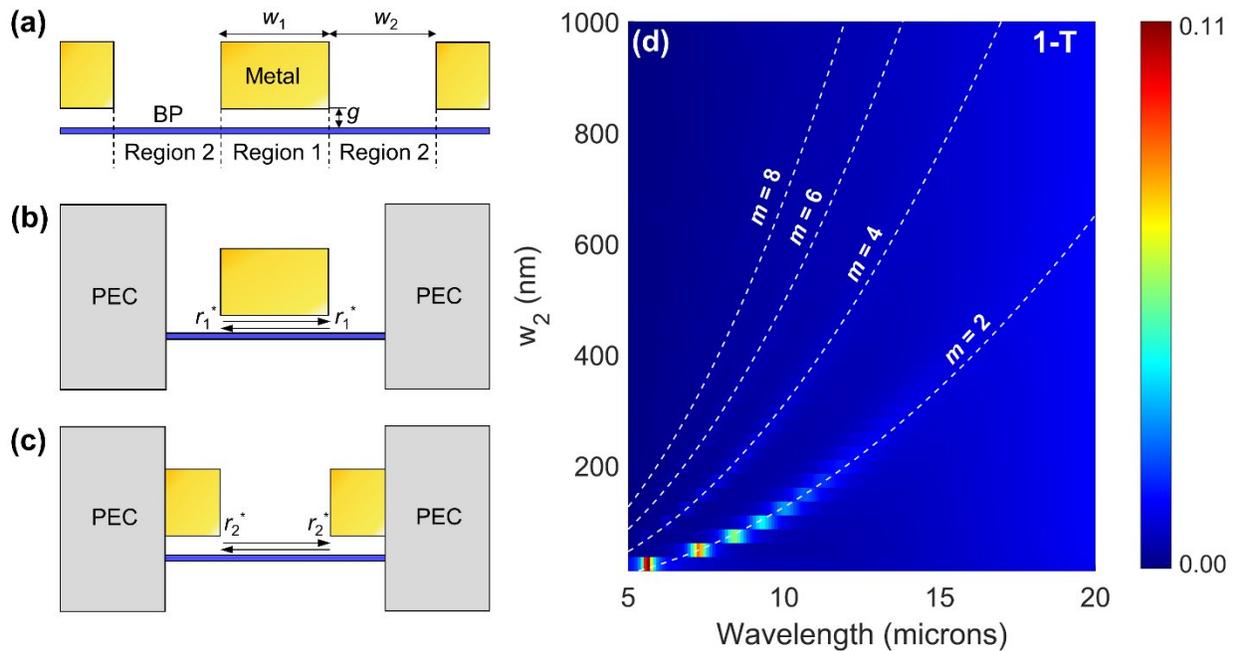

**Figure S3.** (a) Schematic of an acoustic plasmon resonator having a continuous BP sheet on a periodic array of conducting plates. (b) and (c) show two different geometries equivalent with the unit cell in (a). Both ends are terminated by perfect electrical conductors (PEC) to model the periodic structure. $r_1^*$ and $r_2^*$ denote effective reflection coefficients for surface plasmons. (d) Numerical simulations results for extinction as a function of $w_2$. White dashed lines denote the resonant frequencies estimated from a modified Fabry-Perot resonance model.

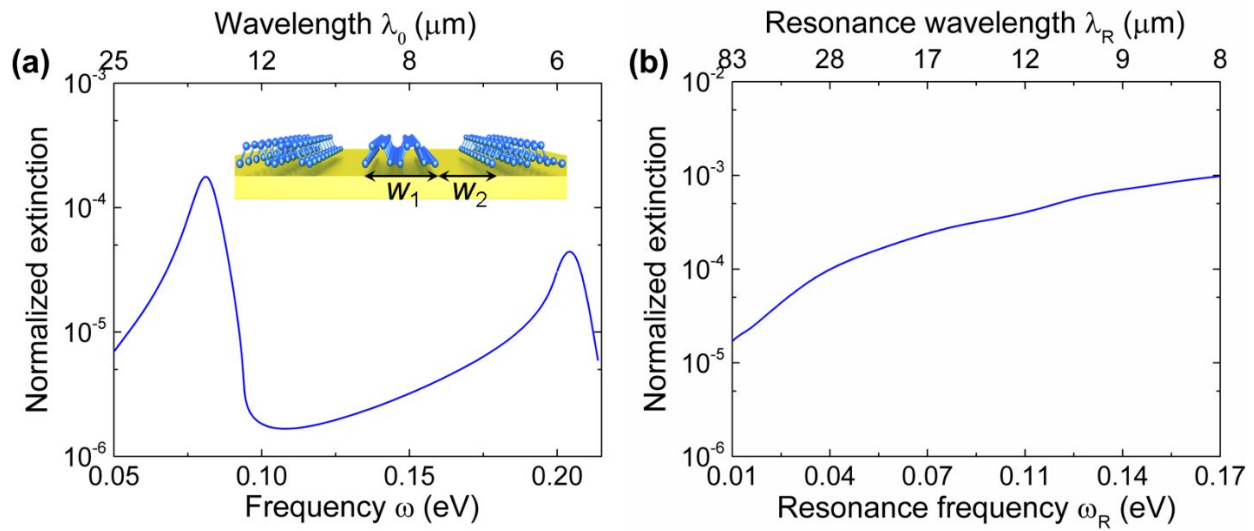

**Figure S4.** (a) Extinction spectra for BP(ribbon)/metal(sheet) case for the ZZ direction, $g = 5$ nm, $w_1 = 61$ nm, and $w_2 = 20$ nm. The zeroth order resonance occurs at $\lambda_0 = 15$ μm. (b) Normalized extinction on resonance as a function of resonance frequency. Here, $w_2$ was fixed to be 20 nm and $w_1$ was varied to shift resonance frequency. Compared to the other designs, the extinction on resonance is negligibly small due to strong reflection form continuous conducting plate.

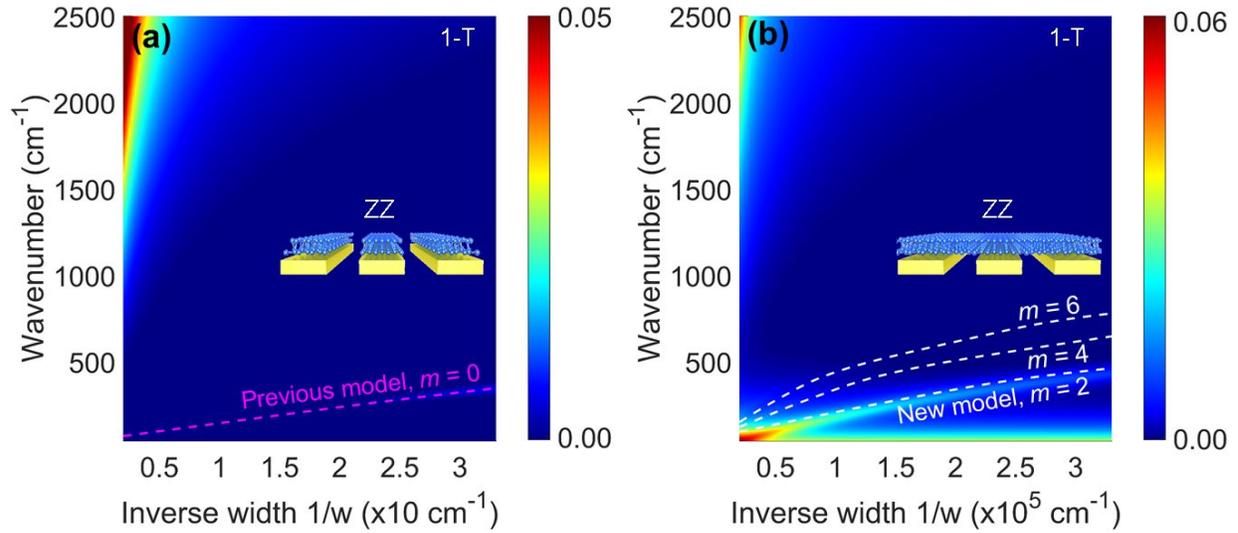

**Figure S5.** Extinction spectra of BP acoustic plasmon resonators for the ZZ direction with (a) periodic BP ribbons and (b) a continuous BP sheet on periodic conducting plates as a function of inverse conducing plate width ($w$) under the illumination of by a normally incident plane wave with transverse magnetic polarization. Magenta dashed lines are the estimated resonant frequencies for different orders of interference ($m$) using the conventional Fabry-Perot resonance equation from the plasmon dispersion and reflection phase obtained previously while white dashed lines come from the modified Fabry-Perot resonance model. Compared to the AC case in Figure 5, the resonances occur at lower frequencies and no resonances are observed at $\omega_{pl}$ (1412 cm$^{-1}$) < $\omega$.

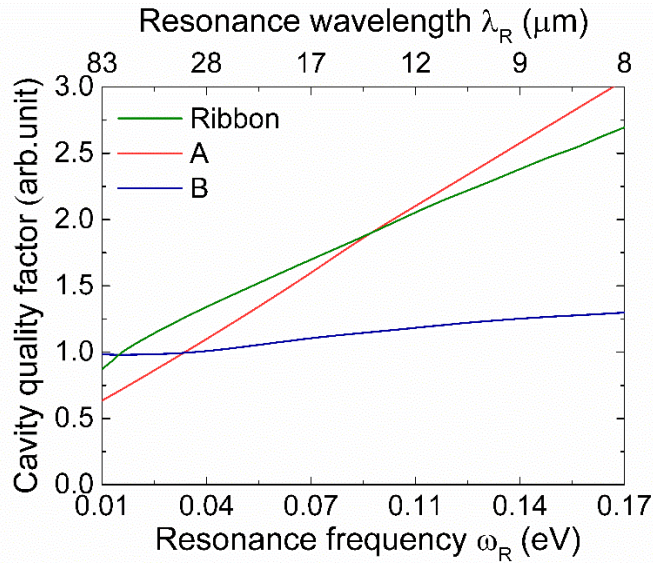

**Figure S6.** Contribution from the quality of cavity to extinction intensity on resonance as a function of resonance frequency $\omega_R$. A and B denote the BP(ribbon)/metal(ribbon) and BP(sheet)/metal(ribbon) resonator, respectively. Here, $g$ was 5 nm and the spacing between two resonator units was fixed to be 20 nm for all cases.